\documentclass[
    ,final            
  ]
  {aipproc}

\layoutstyle{6x9}
\usepackage{amsmath,amssymb}

\newcommand{\lsim}
{\;\raisebox{-.3em}{$\stackrel{\displaystyle <}{\sim}$}\;}

\newcommand{\gmt}{$(g-2)_\mu$}
\newcommand{\br}{{\rm BR}}
\newcommand{\bsg}{BR($b \to s \gamma$)}
\newcommand{\btn}{BR($B_u \to \tau \nu_\tau$)}
\newcommand{\bmm}{BR($B_s \to \mu^+\mu^-$)}

\newcommand{\Och}{\ensuremath{\Omega_\chi h^2}}

\newcommand{\MZ}{M_Z}
\newcommand{\Mh}{M_h}
\newcommand{\MA}{M_A}
\newcommand{\MH}{M_H}
\newcommand{\mt}{m_t}
\newcommand{\mgl}{m_{\tilde g}}

\newcommand{\neu}[1]{\tilde \chi^0_{#1}}

\newcommand{\mste}{m_{\tilde t_1}}
\newcommand{\mstaue}{m_{\staue}}
\newcommand{\staue}{\tilde \tau_1}
\newcommand{\tb}{\tan\beta}
\newcommand{\ecm}{\sqrt{s}}
\newcommand{\tev}{\,\, \mathrm{TeV}}
\newcommand{\gev}{\,\, \mathrm{GeV}}

\graphicspath{{figs/}}

\begin{document}

\thispagestyle{empty}
\setcounter{page}{0}
\def\thefootnote{\fnsymbol{footnote}}

\begin{flushright}
\mbox{}
arXiv:0909.4662 [hep-ph]
\end{flushright}

\vspace{1cm}

\begin{center}

{\fontsize{15}{1} 
\sc {\bf SUSY Predictions for the LHC}}
\footnote{plenary talk given at the {\em BSM--LHC (SUSY\,09)}, 
June 2009, Boston, USA}

\vspace{1cm}

{\sc 
S.~Heinemeyer
\footnote{
email: Sven.Heinemeyer@cern.ch
}
}

\vspace*{1cm}

Instituto de F\'isica de Cantabria (CSIC-UC), Santander,  Spain

\end{center}

\vspace*{0.2cm}

\begin{center} {\bf Abstract} \end{center}
On the basis of frequentist analyses of experimental constraints
from electroweak precision data, \gmt, $B$ physics and cosmological data, we
predict the masses of Higgs bosons and SUSY particles 
of the constrained MSSM (CMSSM) 
with universal soft supersymmetry-breaking mass parameters, and a model
with common non-universal Higgs masses (NUHM1). 
In the CMSSM we find preferences for
sparticle masses that are relatively light.
In the NUHM1 the best-fit values for many sparticle masses are
even slightly smaller, but with greater uncertainties. We
find that at the 95\% C.L.\ all colored particles are in the reach of the LHC.
While the light Higgs boson is bounded from above by $\Mh \lsim 125 \gev$, the
heavy Higgs bosons could well escape the LHC searches, but might be accessible
at the ILC.

\def\thefootnote{\arabic{footnote}}
\setcounter{footnote}{0}

\newpage

\title{SUSY Predictions for the LHC}


\author{S.~Heinemeyer}{
  address={Instituto de F\'isica de Cantabria (CSIC-UC), Santander, Spain}
}

\begin{abstract}
On the basis of frequentist analyses of experimental constraints
from electroweak precision data, \gmt, $B$ physics and cosmological data, we
predict the masses of Higgs bosons and SUSY particles 
of the constrained MSSM (CMSSM) 
with universal soft supersymmetry-breaking mass parameters, and a model
with common non-universal Higgs masses (NUHM1). 
In the CMSSM we find preferences for
sparticle masses that are relatively light.
In the NUHM1 the best-fit values for many sparticle masses are
even slightly smaller, but with greater uncertainties. We
find that at the 95\% C.L.\ all colored particles are in the reach of the LHC.
While the light Higgs boson is bounded from above by $\Mh \lsim 125 \gev$, the
heavy Higgs bosons could well escape the LHC searches, but might be accessible
at the ILC.

\end{abstract}

\maketitle


\section{Introduction}

Supersymmetry (SUSY)~\cite{Nilles:1983ge,Haber:1984rc,Barbieri:1982eh}
is one of the favored ideas for physics beyond
the Standard Model (SM) that may soon be explored at the Large Hadron
Collider (LHC). In several recent papers~\cite{Master1,Master2,Master3},
we presented results from frequentist analyses of 
the parameter spaces of the constrained minimal supersymmetric
extension of the Standard Model (CMSSM) --- in which the
soft supersymmetry-breaking scalar and gaugino masses are each constrained
to universal values $m_0$ and $m_{1/2}$, respectively
(see \cite{Master3} for a comprehensive list of references)
 --- and the NUHM1 --- in which the soft supersymmetry-breaking
contributions to the Higgs masses are allowed a different but common value
(see \cite{Master3} for a comprehensive list of references).
Other statistical analyses in these models can be found in 
~\cite{deBoer:2003xm,Belanger:2004ag,Ellis:2003si,Ellis:2004tc,Ellis:2005tu,Ellis:2006ix,Ellis:2007aa,Ellis:2007ka,Ellis:2007ss,Heinemeyer:2008fb,Bechtle:2004pc,Lafaye:2007vs} and Markov Chain Monte Carlo (MCMC) analyses in 
\cite{Baltz:2004aw,Allanach:2005kz,Allanach:2006jc,Allanach:2006cc,Allanach:2007qj,Allanach:2007qk,Allanach:2008iq,Feroz:2008wr,deAustri:2006pe,Roszkowski:2006mi,Roszkowski:2007fd,Roszkowski:2007va,Trotta:2008bp,Master1,othermod,nonunivgaugino,fitmssm19,fittinoMC}.

Here we review the 
results presented in~\cite{Master3}. They include the parameters of the
best-fit points in the CMSSM and the NUHM1, as well as the 68 and
95\%~C.L.\ regions applyingthe phenomenological, experimental
and cosmological constraints. These include precision electroweak data,
the anomalous magnetic moment of the muon, \gmt, 
$B$-physics observables (the rates for \bsg\ and \btn,
$B_s$ mixing, and the upper limit on \bmm), the bound on the lightest MSSM Higgs
  boson mass, $\Mh$, and the cold dark matter (CDM) density
inferred from astrophysical and cosmological data, assuming that this is
dominated by the relic density of the lightest neutralino, $\Och$.
In~\cite{Master2} we also discussed the sensitivities of the areas of
the preferred regions to changes in the ways in which the major
constraints are implemented. 
We found that the smallest sensitivity was to the CDM density,
and the greatest sensitivity was that to \gmt.


\section{Description of our Frequentist approach}

We define a global $\chi^2$ likelihood function, which combines all
theoretical predictions with experimental constraints:
\begin{align}
\chi^2 &= \sum^N_i \frac{(C_i - P_i)^2}{\sigma(C_i)^2 + \sigma(P_i)^2}
+ {\chi^2(\Mh) + \chi^2(\br(B_s \to \mu\mu))}
\nonumber \\[.2em]
&+ {\chi^2(\mbox{SUSY search limits})}
+ \sum^M_i \frac{(f^{\rm obs}_{{\rm SM}_i}
              - f^{\rm fit}_{{\rm SM}_i})^2}{\sigma(f_{{\rm SM}_i})^2}
\label{eqn:chi2}
\end{align} 
Here $N$ is the number of observables studied, $C_i$ represents an
experimentally measured value (constraint) and each $P_i$ defines a
prediction for the corresponding constraint that depends on the
supersymmetric parameters.
The experimental uncertainty, $\sigma(C_i)$, of each measurement is
taken to be both statistically and systematically independent of the
corresponding theoretical uncertainty, $\sigma(P_i)$, in its
prediction. We denote by
$\chi^2(\Mh)$ and $\chi^2(\br(B_s \to \mu\mu))$ the $\chi^2$
contributions from the two measurements for which only one-sided
bounds are available so far, as discussed below.
Furthermore we include the lower limits from the direct searches
for SUSY particles at LEP~\cite{LEPSUSY} as one-sided limits, denoted by 
``$\chi^2(\mbox{SUSY search limits})$'' in eq.~(\ref{eqn:chi2}).

We stress that in \cite{Master3} (as in~\cite{Master1,Master2})
the three SM parameters
$f_{\rm SM} = \{\Delta\alpha_{\rm had}, \mt, \MZ \}$ are included as fit
parameters and allowed to vary with their current experimental
resolutions $\sigma(f_{\rm SM})$. We do not
include $\alpha_s$ as a fit parameter, 
which would have only a minor impact on the analysis.

Formulating the fit in this fashion has the advantage that the
$\chi^2$ probability, $P(\chi^2, N_{\rm dof})$,
properly accounts for the number of degrees of freedom, $N_{\rm dof}$,
in the fit and thus represents a quantitative and meaningful measure for
the ``goodness-of-fit.'' In previous studies \cite{Master1},
$P(\chi^2, N_{\rm dof})$ has been verified to have a flat distribution,
thus yielding a reliable estimate of the confidence level for any particular
point in parameter space. Further, an important aspect of the
formulation is that all model parameters are varied simultaneously in
our MCMC sampling, and care is exercised to fully explore the
multi-dimensional space, including possible interdependencies between
parameters.  All confidence levels for selected model parameters are
performed  by scanning over the desired parameters while  
minimizing the $\chi^2$ function with respect to all other model parameters. 
The function values where $\chi^2(x)$ is found to be equal to 
$\chi^2_{min}+ \Delta \chi^2$ determine the confidence level
contour. For two-dimensional parameter scans we use 
$\Delta \chi^2 =2.28 (5.99)$ to determine the 68\%(95\%) confidence
level contours. 
Only experimental constraints are imposed when deriving confidence level
contours, without any arbitrary or direct constraints placed on model
parameters themselves.
This leads to robust and statistically meaningful
estimates of the total 68\% and 95\% confidence levels,
which may be composed of multiple separated contours.

The experimental constraints used in our analyses are listed in
Table~1 in \cite{Master3}. 
One important comment concerns our implementation of the LEP
constraint on $\Mh$. The value quoted in the Table, $\MH > 114.4 \gev$,
was derived within the SM~\cite{Barate:2003sz}, and is applicable to the
CMSSM, in which 
the relevant Higgs couplings are very similar to those in the 
SM~\cite{Ellis:2001qv,Ambrosanio:2001xb}, so that the SM
exclusion results can be used, supplemented with an additional theoretical
uncertainty:
we evaluate the $\chi^2(\Mh)$ contribution within the CMSSM using the
formula
\begin{align}
\chi^2(\Mh) = \frac{(\Mh - \Mh^{\rm limit})^2}{(1.1 \gev)^2 + (1.5 \gev)^2}~,
\label{chi2Mh}
\end{align}
with $\Mh^{\rm limit} = 115.0 \gev$ 
for $\Mh < 115.0 \gev$. 
Larger masses do not receive a $\chi^2(\Mh)$ contribution. 
We use $115.0 \gev$ so as to incorporate
a conservative consideration of experimental systematic effects.
The $1.5 \gev$ in the denominator corresponds to a convolution of
 the likelihood function with a Gaussian function, $\tilde\Phi_{1.5}(x)$,
normalized to unity and centered around $\Mh$, whose width is $1.5 \gev$,
representing the theory uncertainty on $\Mh$~\cite{Degrassi:2002fi}.
In this way, a theoretical uncertainty of up to $3 \gev$ is assigned for 
$\sim 95\%$ of all $\Mh$ values corresponding to one CMSSM parameter point. 
The $1.1 \gev$ term in the denominator corresponds to a parametrization
of the $CL_s$ curve given in the final SM LEP Higgs
result~\cite{Barate:2003sz}. 

Within the NUHM1 the situation is somewhat more involved, since, for
instance, a strong suppression of the $ZZh$ coupling can occur,
invalidating the SM exclusion bounds. 
In order to find a more reliable 95\% C.L.\ exclusion limit for $\Mh$ in the
case that the SM limit cannot be applied, we use the following procedure.
The main exclusion bound from LEP searches comes from the channel
$e^+e^- \to ZH, H \to b \bar b$. The Higgs boson mass limit in this
channel is given as a function of the $ZZH$ coupling in~\cite{Schael:2006cr}. 
A reduction in the $ZZh$ coupling in the
NUHM1 relative to its SM value can be translated into a lower
limit on the lightest NUHM1 Higgs mass, $\Mh^{{\rm limit},0}$, shifted
to lower values with respect to the SM limit of $114.4 \gev$. (The
actual number is obtained using the code {\tt HiggsBounds}~\cite{higgsbounds}
that incorporates the LEP (and Tevatron) limits on neutral Higgs boson
searches.) 
For values of $\Mh \lsim 86 \gev$ the reduction of the $ZZh$
couplings required to evade the LEP bounds becomes very strong, and we
add a brick-wall contribution to the $\chi^2$ function below this value
(which has no influence on our results).
Finally, eq.~(\ref{chi2Mh}) is used with 
$\Mh^{\rm limit} = \Mh^{{\rm limit},0} + 0.6 \gev$ to ensure a smooth
transition to the SM case, see~\cite{Master3} for more details.

The numerical evaluation of the frequentist likelihood function
using the constraints has been performed with the 
{\tt MasterCode}~\cite{Master1,Master2,Master3},
which includes the following theoretical codes. For the RGE running of
the soft SUSY-breaking parameters, it uses
{\tt SoftSUSY}~\cite{Allanach:2001kg}, which is combined consistently
with the codes used for the various low-energy observables. 
At the electroweak scale we have included various codes:
{\tt FeynHiggs}~\cite{Degrassi:2002fi,Heinemeyer:1998np,Heinemeyer:1998yj,Frank:2006yh}  
is used for the evaluation of the Higgs masses and
$a_\mu^{\rm SUSY}$ (see also
\cite{Moroi:1995yh,Degrassi:1998es,Heinemeyer:2003dq,Heinemeyer:2004yq}).
For the difference between the SM and the experimental value we used
$\Delta a_\mu = (30.2 \pm 8.8) \times 10^{-10}$~\cite{Davier:2007ua}
based on $e^+e^-$ data,
see also~\cite{Miller:2007kk,Jegerlehner:2007xe,Passera:2008jk}. 
A new evaluation, including new {\it B{\tiny A}B{\tiny AR}} data, yields
$\Delta a_\mu = (24.6 \pm 8.0) \times 10^{-10}$~\cite{Davier:2009zi}.
Using this value could have a small impact on our results.
We note that recently a new 
$\tau$~based analysis has appeared~\cite{g-2taunew}, which yields 
a $\sim 1.9\,\sigma$ deviation from the SM prediction. 
For flavor-related observables we use 
{\tt SuFla}~\cite{Isidori:2006pk,Isidori:2007jw} as well as 
{\tt SuperIso}~\cite{Mahmoudi:2008tp,Eriksson:2008cx}, and
for the electroweak precision data we have included 
a code based on~\cite{Heinemeyer:2006px,Heinemeyer:2007bw}.
Finally, for dark-matter-related observables, 
{\tt MicrOMEGAs}~\cite{Belanger:2006is,Belanger:2001fz,Belanger:2004yn} and
{\tt DarkSUSY}~\cite{Gondolo:2005we,Gondolo:2004sc} 
have been used.
We made extensive use of the SUSY Les Houches
Accord~\cite{Skands:2003cj,Allanach:2008qq} 
in the combination of the various codes within the {\tt MasterCode}.


\section{Results for Sparticle Masses}

For the parameters of the best-fit CMSSM point we find
$m_0 = 60 \gev$,  $m_{1/2} = 310 \gev$,  $A_0 = 130 \gev$, $\tb = 11$
and $\mu = 400 \gev$, 
yielding the overall $\chi^2/{\rm N_{\rm dof}} = 20.6/19$ (36\% probability) 
and nominally $\Mh = 114.2 \gev$.
The corresponding parameters of
the best-fit NUHM1 point are $m_0 = 150 \gev$, $m_{1/2} = 270 \gev$,
$A_0 = -1300 \gev$, $\tb = 11$ and
$m_{h_1}^2  = m_{h_2}^2 = - 1.2 \times 10^6 \gev^2$ or, equivalently,
$\mu = 1140 \gev$, yielding
$\chi^2 = 18.4$ (corresponding to a similar fit probability to the CMSSM)
and $\Mh = 120.7 \gev$. 

\begin{figure*}[htb!]
\resizebox{8cm}{!}{\includegraphics{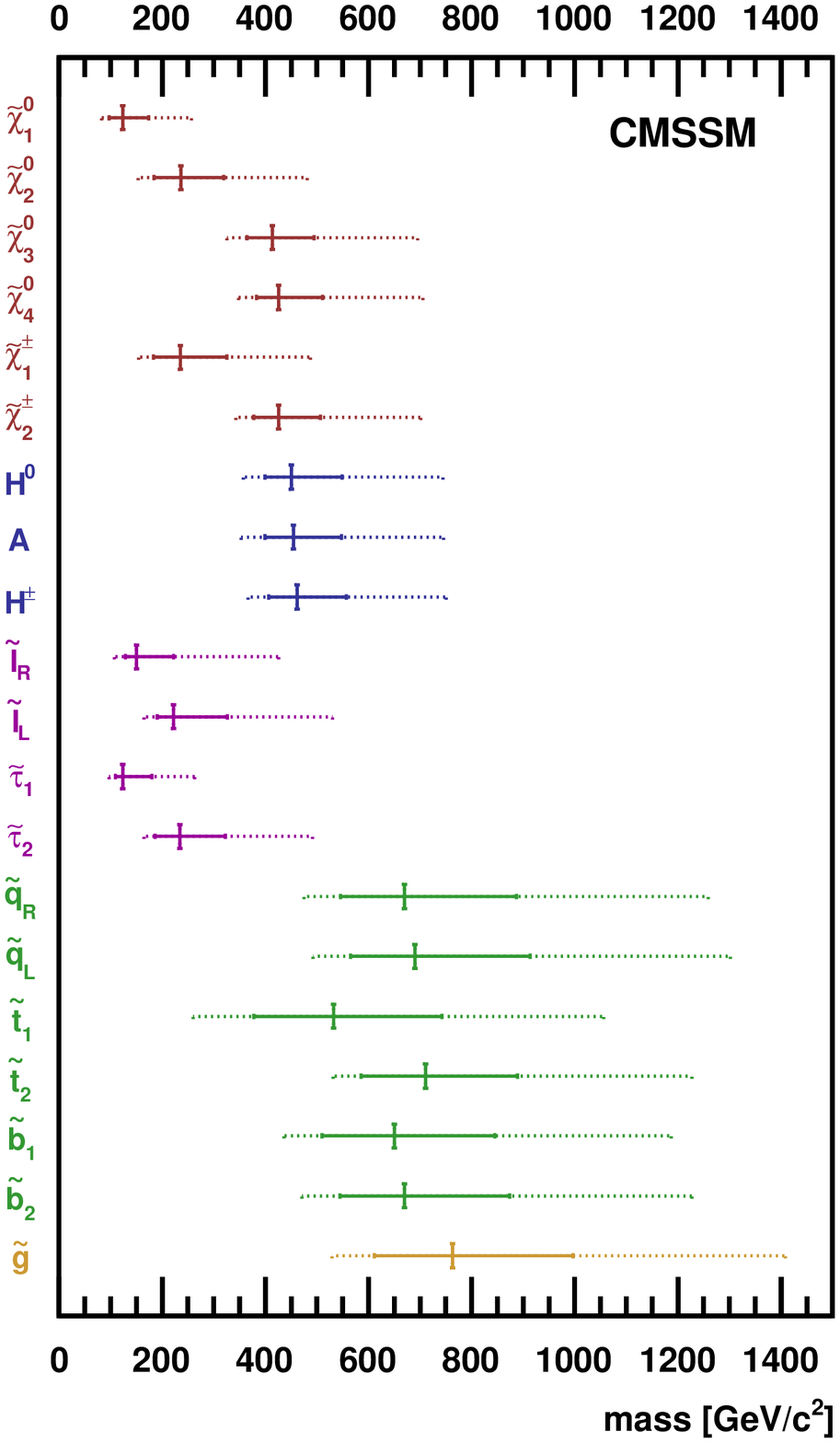}}
\resizebox{8cm}{!}{\includegraphics{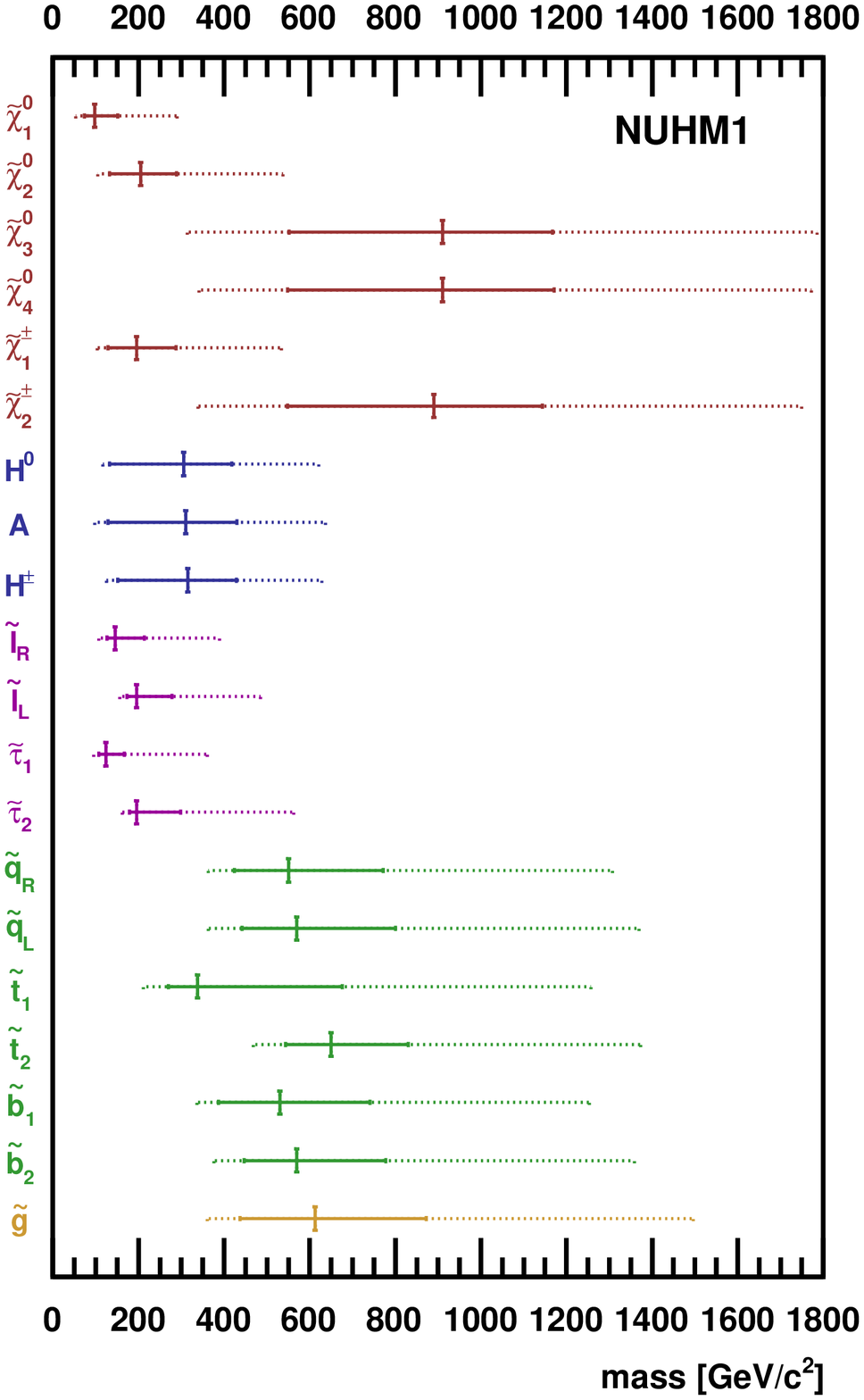}}
\caption{\it Spectra in the CMSSM (left) and the NUHM1
(right)~\cite{Master3}. The vertical 
solid lines indicate the best-fit values, the horizontal solid lines
are the 68\% C.L.\
ranges, and the horizontal dashed lines are the 95\% C.L.\ ranges for the
indicated mass parameters. 
}
\label{fig:spectrum}
\end{figure*}

We now review the results for the predictions of sparticles masses in the
CMSSM and the NUHM1, 
which are summarized in Fig.~\ref{fig:spectrum}. The results for the CMSSM
spectrum are shown in the left plot, and for the NUHM1 in the right plot. 
We start our discussion with the gluino mass, $\mgl$.
In both the CMSSM and the NUHM1, the
best-fit points have relatively low values of $\mgl \sim 750$ and $\sim
600 \gev$, respectively. These favored values are well within the range
even of the early operations of the LHC with reduced centre-of-mass
energy and limited luminosity. However, even quite large values of 
$\mgl \lsim 2.5 \tev$
are allowed at the 3-$\sigma$ ($\Delta \chi^2 = 9$) level 
(not shown in Fig.~\ref{fig:spectrum}). The LHC
should be  able to discover a gluino with $\mgl \sim 2.5 \tev$ with
100/fb of integrated luminosity at 
$\ecm = 14 \tev$~\cite{atlastdr,cmstdr}, and the proposed 
SLHC luminosity upgrade to 1000/fb of integrated luminosity at 
$\ecm = 14 \tev$ should permit the discovery of a gluino with 
$\mgl \sim 3 \tev$~ \cite{Gianotti:2002xx}. 
However, Fig.~\ref{fig:spectrum} does demonstrate that, whilst there are good
prospects for discovering SUSY in early LHC running~\cite{Master2}, 
this cannot be `guaranteed'.

The central values of the masses of the supersymmetric
partners of the $u, d, s, c, b$ quarks are slightly lighter than the
gluino, as seen in 
Fig.~\ref{fig:spectrum}. The difference between the gluino and the
squark masses is sensitive primarily to $m_0$.
The reason is that the preferred regions of the
parameter space in both the CMSSM
and the NUHM1 are in the $\neu{1}$-slepton coannihilation
region~\cite{Master2,Master3} where $m_0 < m_{1/2}$. Here $m_0$ makes only
small contributions to the central values of the
squark masses. The SUSY partners of the left-handed components of
the four lightest quarks, the ${\tilde q_L}$, 
are predicted to be slightly heavier than the corresponding right-handed
squarks, ${\tilde q_R}$, as seen by comparing the mass ranges in
Fig.~\ref{fig:spectrum}. As in the case of the gluino,
squark masses up to $\sim 2.5 \tev$ are allowed at the 3-$\sigma$
level. Comparing 
the left and right panels, we see that the squarks are predicted to be
somewhat lighter in the NUHM1 than in the CMSSM, 
but this difference is small compared
with the widths of the corresponding likelihood functions.

Turning now to the likelihood functions for the mass of the lighter stop, 
$\mste$, we find that it is shifted to values somewhat lower than for
the other squark flavors. It can also be seen
that the 2-$\sigma$ range of its likelihood function differ from those
of the gluino and the other squarks, reflecting the importance of scalar
top mixing. We recall that this depends strongly on the trilinear
soft SUSY-breaking parameter $A_t$ and the
Higgs mixing parameter $\mu$, as well as on the precise value
of $\mt$. 

In the case of the lighter stau $\staue$, see its range in
Fig.~\ref{fig:spectrum}, the mass is very similar to
that of the LSP $\neu{1}$ in the coannihilation region, but this is not
the case in the rapid-annihilation $H, A$ funnel region,
see~\cite{Master3} for details.
In the case of the NUHM1 rapid annihilation is possible also for low
$\tb$, leading to larger values of $m_0$ than in the CMSSM also for
relatively small values of $\mstaue$.

The scalar taus as well as the other scalar leptons are expected to be
relatively light, as can be seen in Fig.~\ref{fig:spectrum}. They would
partially be in the reach of the ILC(500) (i.e.\ with $\sqrt{s} = 500 \gev$)
and at the 95\% C.L.\ nearly all be in the reach of the
ILC(1000)~\cite{teslatdr,Brau:2007zza}. 
This also holds for the two lighter neutralinos and the light chargino.


\section{Prediction of Higgs Boson Masses}

\begin{figure*}[htb!]
\resizebox{8cm}{!}{\includegraphics{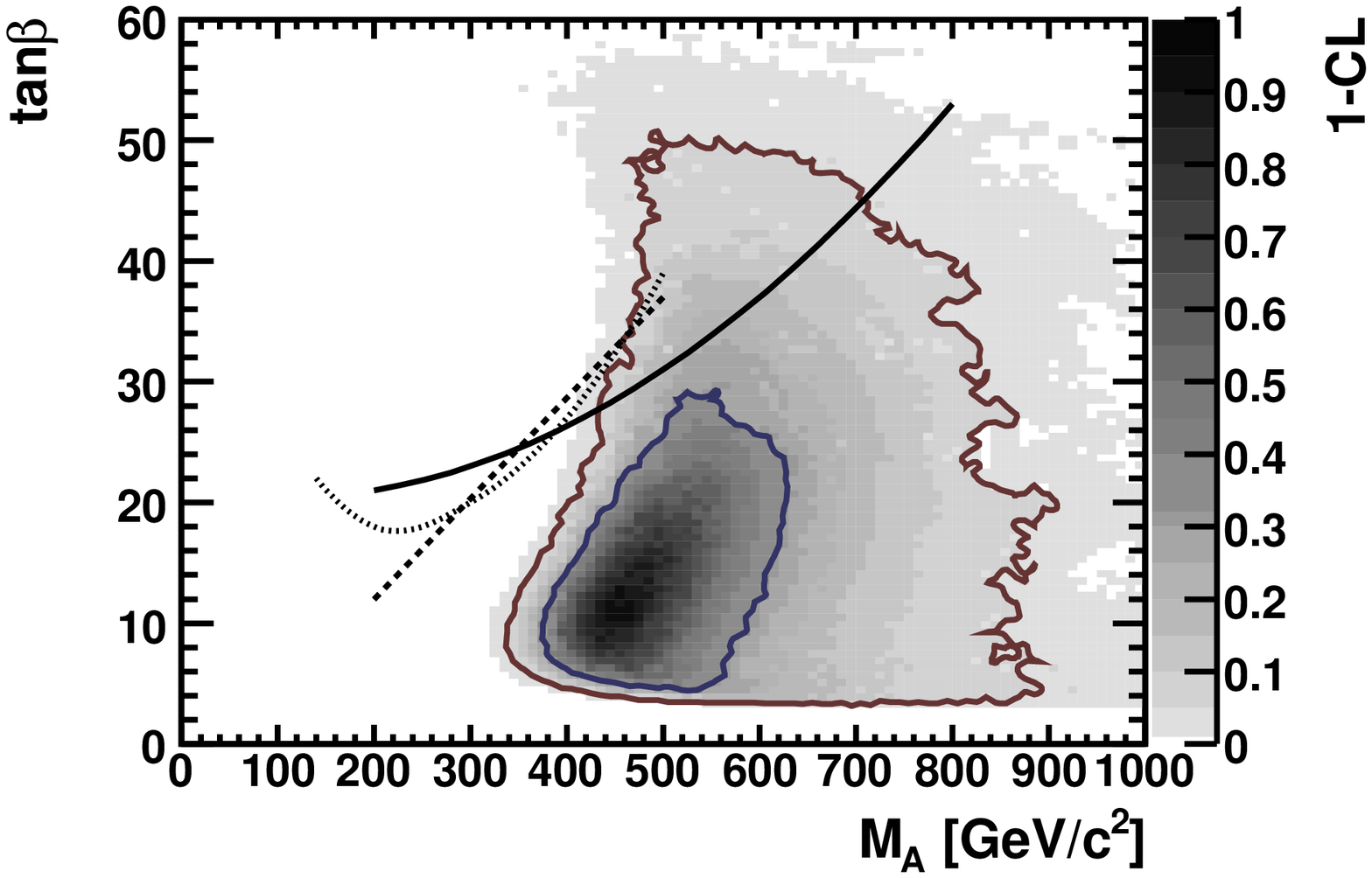}}
\resizebox{8cm}{!}{\includegraphics{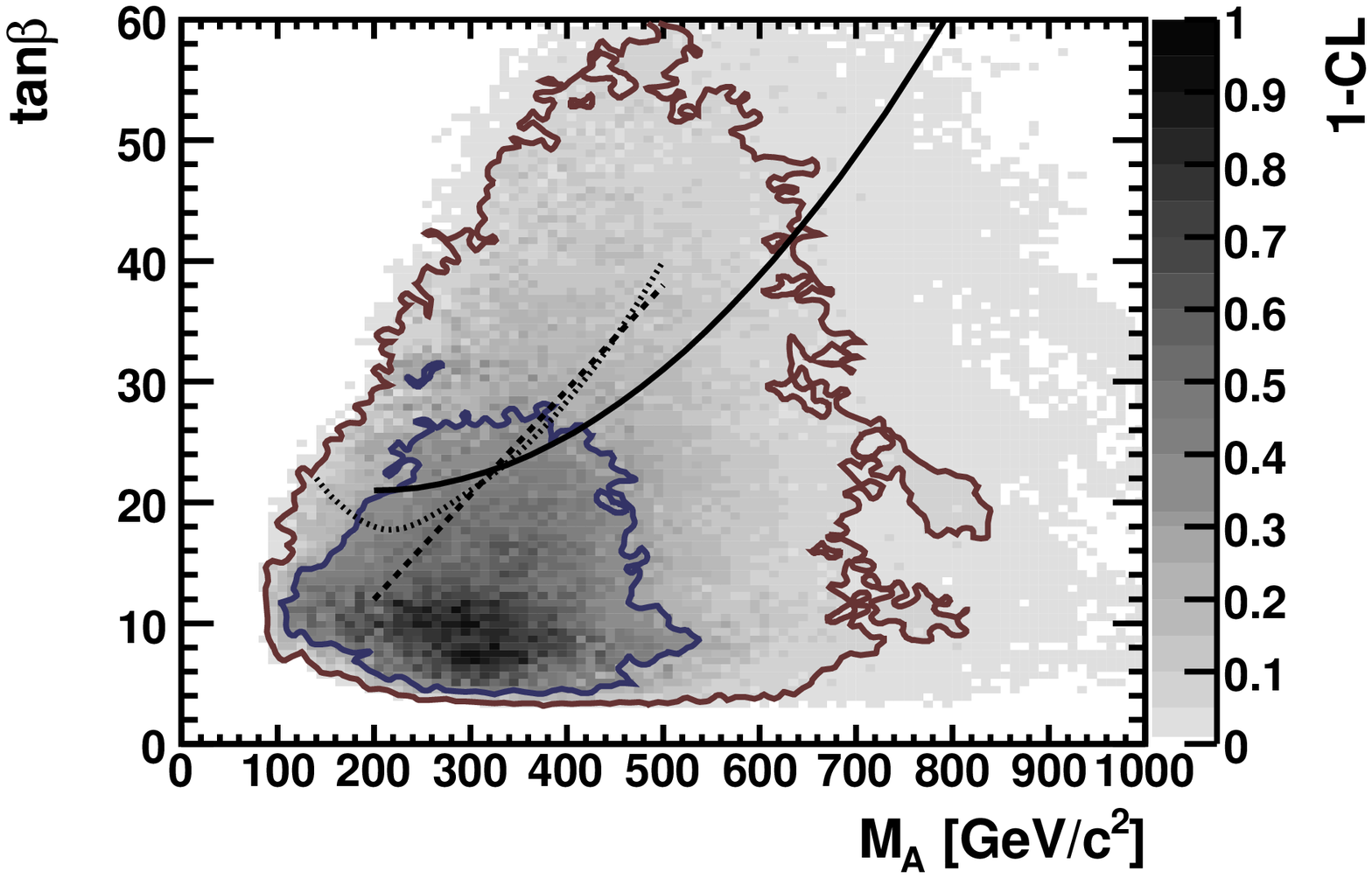}}
\vspace{-2em}
\caption{\it The correlations between $\MA$ and $\tb$
in the CMSSM (left panel) and in the NUHM1 (right panel)~\cite{Master3}.
Also shown are the 5-$\sigma$ discovery contours 
for observing the heavy MSSM Higgs bosons $H, A$ 
in the three decay channels
$H,A \to \tau^+\tau^- \to {\rm jets}$ (solid line), 
$\rm{jet}+\mu$ (dashed line), $\rm{jet}+e$ (dotted line)
at the LHC. The discovery contours have been obtained using an 
analysis that assumed 30 or 60~fb$^{-1}$ 
collected with the CMS detector~\cite{cmstdr,cmsHiggs}.
}
\label{fig:mAtb}
\vspace{3em}
\end{figure*}

In Fig.~\ref{fig:mAtb} we display the favored regions in the
$(\MA, \tb)$ planes 
for the CMSSM and NUHM1. We see that they are broadly similar, with little
correlation between the two parameters.
Concerning $\tb$, one can observe that while the best fit values lie at
$\tb \approx 11$, the 68 (95)\% C.L.\ areas reach up 30 
$\tb \approx 30 (50$-$60)$. 
The existing Higgs discovery analyses (performed in the various
benchmark scenarios~\cite{Ellis:2007aa,Ellis:2007ka,benchmark2,benchmark3}) 
cannot directly be applied to the $(\MA, \tb)$ planes in
Fig.~\ref{fig:mAtb}. In order to assess the prospects for discovering
heavy Higgs bosons at the LHC in this context,
we follow the analysis in~\cite{cmsHiggs}, which assumed 30
or 60~fb$^{-1}$ collected with the CMS detector. For evaluating the
Higgs-sector observables including higher-order corrections we use 
the soft
SUSY-breaking parameters of the best-fit points in the CMSSM and the
NUHM1, respectively. We show in Fig.~\ref{fig:mAtb} the 5-$\sigma$
discovery contours for the three decay channels 
$H,A \to \tau^+\tau^- \to {\rm jets}$ (solid lines), $\rm{jet}+\mu$ (dashed
lines) and $\rm{jet}+e$ (dotted lines).
The parameter regions above and to the left of the curves are within reach 
of the LHC with about 30~fb$^{-1}$ of integrated luminosity.
We see that most of the highest-CL regions lie beyond this reach,
particularly in the CMSSM.
At the ILC(1000) masses up to $\MA \lsim 500 \gev$ can be probed. 
Within the CMSSM this includes the best-fit point, and within the NUHM1
nearly the whole 68\% C.L.\ area can be covered.

\begin{figure*}[htb!]
\resizebox{8cm}{!}{\includegraphics{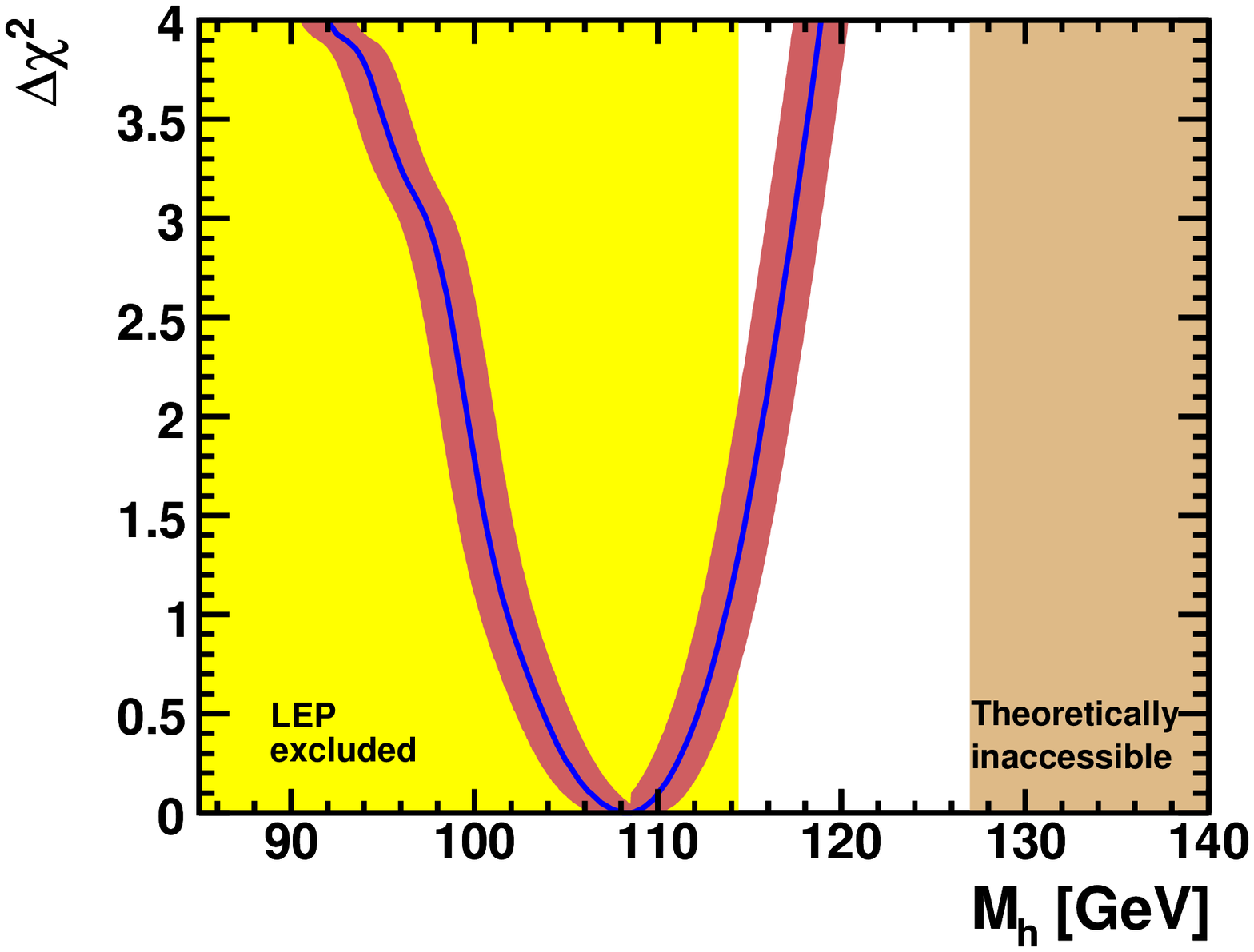}}
\resizebox{8cm}{!}{\includegraphics{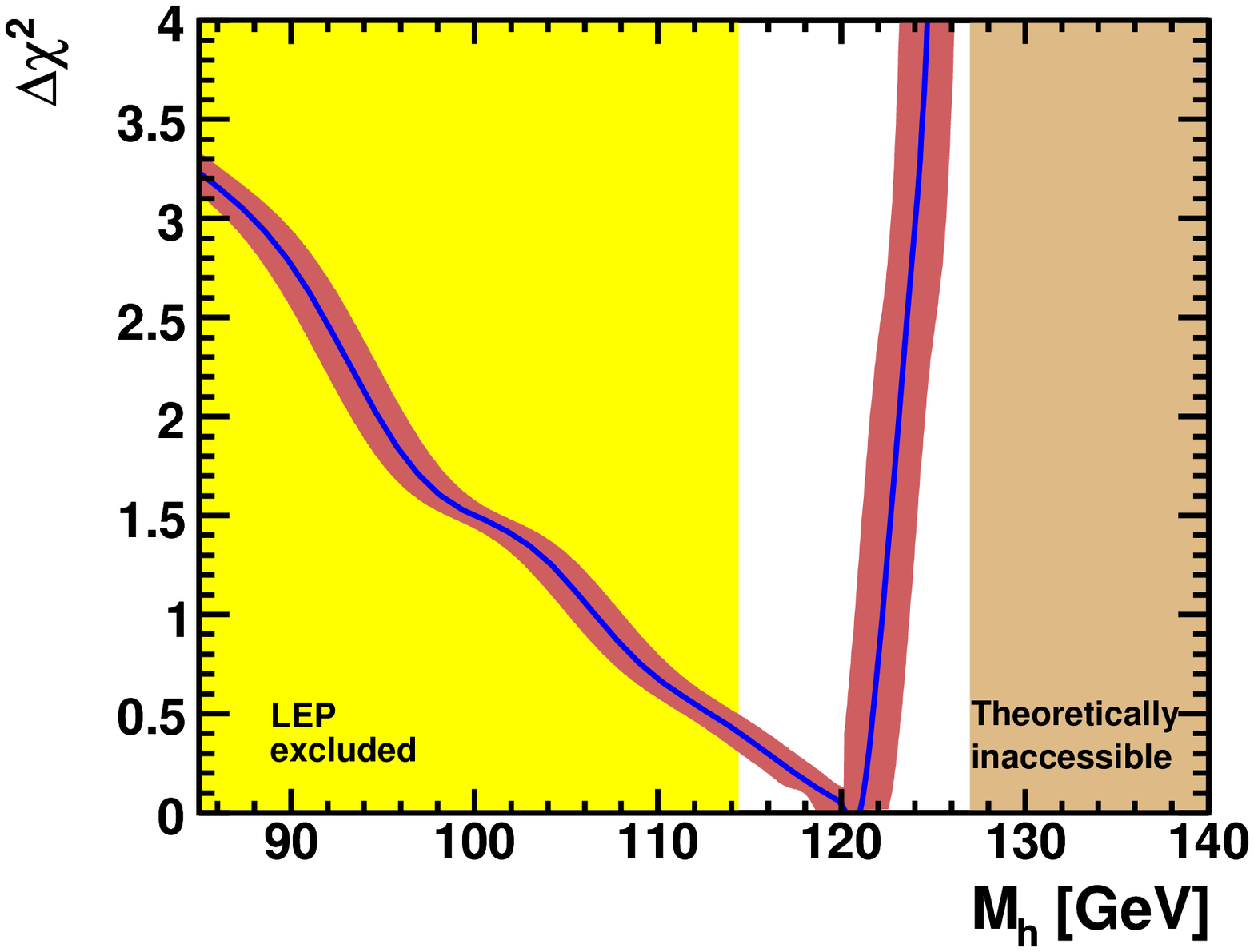}}
\vspace{-1cm}
\caption{\it The $\chi^2$ functions for $\Mh$ in the CMSSM (left) and
  the NUHM1 (right)~\cite{Master3}, 
including the theoretical uncertainties (red bands). Also shown is the mass
range excluded for a SM-like Higgs boson (yellow shading),
and the ranges theoretically inaccessible in the supersymmetric models
studied. 
}
\label{fig:redband}
\end{figure*}

Finally we discuss 
the likelihood functions for $\Mh$ within the CMSSM and NUHM1 
frameworks obtained {\em when dropping} the
contribution to $\chi^2$ from the direct Higgs searches at LEP.
The results are 
shown in the left and right panels of Fig.~\ref{fig:redband}, respectively. 
The left plot updates that for the CMSSM given in~\cite{Master1}.

It is well
known that the central value of the Higgs mass in a SM
fit to the precision electroweak data lies below
100~GeV~\cite{ewpoMoriond2009,lepewwg}, 
but the theoretical (blue band) and experimental uncertainties 
in the SM fit are such that they are still compatible at the 
95\% C.L.\ 
with the direct lower limit of
114.4~GeV~\cite{Barate:2003sz} derived 
from searches at LEP. In the case of the CMSSM and NUHM1,
one may predict $\Mh$ on the basis of the underlying model
parameters, with a 1-$\sigma$ uncertainty of 1.5~GeV~\cite{Degrassi:2002fi},
shown as a red band in Fig.~\ref{fig:redband}. Also shown in
Fig.~\ref{fig:redband} are the LEP exclusion on a SM Higgs
(yellow shading)
and the ranges that are theoretically inaccessible in the
supersymmetric models studied (beige shading). 
The LEP exclusion is directly applicable to the CMSSM,
since the $h$ couplings are essentially indistinguishable from
those of the SM Higgs boson~\cite{Ellis:2001qv,Ambrosanio:2001xb}, but
this is not necessarily the case in the NUHM1, as discussed earlier.

In the case of the CMSSM, we see in the left panel of
Fig.~\ref{fig:redband} that the minimum of the $\chi^2$
function occurs below the LEP exclusion limit. 
While the tension between the $\chi^2$ function for $\Mh$ 
arising from the CMSSM
fit and the LEP exclusion limit has slightly increased compared to the
earlier analysis performed in \cite{Master1}, the fit result is still
compatible at the 95\% C.L.\ with the search limit, 
similarly to the SM case. As we found in~\cite{Master3} 
a global fit including the LEP constraint has acceptable $\chi^2$.
In the case of the NUHM1, shown in the right panel of
Fig.~\ref{fig:redband}, we see that the minimum of the $\chi^2$
function occurs {\it above} the LEP lower limit on the mass of a SM 
Higgs. Thus, 
within the NUHM1 the combination of all other experimental
constraints {\em naturally} evades the LEP Higgs constraints, and no
tension between $\Mh$ and the experimental bounds exists.



\begin{theacknowledgments}
We thank O.~Buchm\"uller, R.~Cavanaugh, A.~De~Roeck, J.~Ellis,
H.~Fl\"acher, G.~Isidori, K.~Olive, F.~Ronga and G.~Weiglein with whom
the results presented here have been obtained.
This work has been supported 
in part by the European Community's Marie-Curie Research
Training Network under contract MRTN-CT-2006-035505
`Tools and Precision Calculations for Physics Discoveries at Colliders'
(HEPTOOLS). 
\end{theacknowledgments}

\bibliographystyle{aipprocl} 

\end{document}